\documentclass[preprint]{aastex}
\usepackage{lscape}
\begin{document}

\title{Colors and Kinematics of L Dwarfs From the Sloan Digital Sky Survey}

\author{Sarah J. Schmidt\altaffilmark{1}, Andrew A. West\altaffilmark{2,3}, Suzanne L.
Hawley\altaffilmark{1}, J. Sebastian Pineda\altaffilmark{2}}

\altaffiltext{1} {Department of Astronomy, University of
Washington, Box 351580, Seattle, WA 98195}
\altaffiltext{2} {Massachusetts Institute of Technology, Kavli Institute for
Astrophysics and Space Research, Building 37, 77 Massachusetts
Avenue, Cambridge, MA 02139, USA}
\altaffiltext{3} {Boston University, Department of Astronomy, 725 Commonwealth Ave, Boston, MA, 02215}

\begin{abstract}
We present a sample of 484 L dwarfs, 210 of which are newly discovered from the Sloan Digital Sky Survey (SDSS) Data Release 7 spectroscopic database. We combine this sample with known L dwarfs to investigate their $izJHK_S$ colors. Our spectroscopically selected sample has $\sim$0.1 magnitude bluer median $J-K_S$ colors at a given spectral type (for L0 to L4) than previously known L dwarfs, which reflects a bias towards redder L dwarfs in past selection criteria. We present photometric distance relations based on $i-z$ and $i-J$ colors and derive distances to our L dwarf sample. We combine the distances with SDSS/2MASS proper motions in order to examine the tangential velocities. For the majority of our spectroscopic sample, we measured radial velocities and present three dimensional kinematics. We also provide H$\alpha$ detections for the fraction of our sample with sufficient quality spectra. Comparison of the velocities of our L dwarf sample to a kinematic model shows evidence for both cold and hot dynamical populations, consistent with young and old disk components. The dispersions of these components are similar to those found for M dwarfs. We also show that $J-K_S$ color is correlated with velocity dispersion, confirming a relationship between $J-K_S$ color and age.
\end{abstract}


\section{Introduction}
\label{sec:intro}

The Sloan Digital Sky Survey \citep[SDSS]{SDSS} is a valuable tool for discovering nearby L dwarfs. Initial work typically employed optical and near-infrared selection criteria combined with follow-up spectroscopy of candidate objects on other telescopes \citep{Fan2000,Schneider2002,Geballe2002}. \citet[hereafter H02]{Hawley2002} was the first study that included SDSS spectroscopy for a large sample. \citet{West2004,West2008} included L dwarfs in their investigations of magnetic activity, but \citet{West2004} used a strict quality cut on the spectra and \citet{West2008} restricted their analysis to L0 and earlier dwarfs. Other work relied on the SDSS photometric database for target selection, but used infrared spectroscopy from other facilities to confirm candidate dwarfs \citep{Knapp2004,Chiu2006}. Recently, the search for additional SDSS dwarfs has focused on cross-matches of SDSS and 2MASS for candidate selection, either taking advantage of additional colors or using proper motion as a selection tool \citep{Metchev2008,Scholz2009,Sheppard2009,Zhang2009}. Since H02, there has been no comprehensive search of the SDSS spectroscopic database to select and confirm L dwarfs. During that time, the SDSS spectroscopic database has grown by roughly an order of magnitude, from $\sim$180,000 sources \citep[DR1;][]{Abazajian2003} to $\sim$1.6 million sources \citep[DR7;][]{Abazajian2009}.

The reliance of many of the past searches on color and proper motion selection criteria may have introduced significant biases into the sample of presently known L dwarfs. The recent discovery of a new L dwarf within 10~pc \citep[SDSS 1416+13;][]{Schmidt2010,Bowler2010} points to a need to use a different set of criteria in order to investigate a representative sample of L dwarfs that spans the entire range of colors and kinematic properties. Selecting objects based on their SDSS spectra allows us to relax the color criteria in order to mitigate the possible biases in the L dwarf sample.

The use of SDSS spectra also allows us to measure radial velocities for a large number of L dwarfs. Previous work has found that the L dwarf population is kinematically younger than most thin disk stars but was limited either by small numbers \citep{ZapateroOsorio2007} or by a lack of radial velocities, restricting kinematic analysis to two dimensions \citep{Schmidt2007,Faherty2009}.  A larger sample with both proper motions and radial velocities is needed to confirm these results and to allow an investigation into the ages of L dwarfs.

While many of the L dwarfs in the SDSS spectroscopic database have relatively low signal to noise ratios (SNR), most have sufficient flux to assign a spectral type and determine the radial velocity. We combine these data with photometric and proper motion information from SDSS and 2MASS to investigate the colors and kinematics of SDSS L dwarfs. In Section~\ref{sec:sample}, we introduce our spectroscopic sample, including a brief discussion of H$\alpha$ emission, and describe our photometric sample which includes both new and known L and T dwarfs. We present colors and derive photometric distance relations in Section~\ref{sec:color}. Section~\ref{sec:kine} describes our kinematic measurements and the resulting velocity distributions.

\section{Sample}
\label{sec:sample}
\subsection{Spectroscopic Sample}
The SDSS \citep{SDSS,Stoughton2002,Pier2003,Ivezic2004} is a multicolor \citep[$ugriz$][]{Fukugita1996,Gunn1998,Hogg2001,Smith2002,Tucker2006}
photometric and spectroscopic survey centered on the northern Galactic cap. The most recent data release \citep[DR7]{Abazajian2009} comprises 11000 deg$^2$ of imaging, yielding photometry of $\sim$357 million unique objects. SDSS also has twin fiber-fed spectrographs which simultaneously obtain 640 medium-resolution (R $\sim$1800), flux-calibrated, optical (3800-9200\AA) spectra per 3$^{\circ}$ plate \citep{Abazajian2004}. While the SDSS primarily targets extragalactic objects, of the $\sim$1.6 million spectra in DR7, $\sim$84,000 are from M or later type stars  \citep{Abazajian2009}. 

Our initial sample of 13,874 spectra was selected from the SDSS photometric and spectroscopic databases by requiring $i-z > 1.4$. This large number of spectra and loose selection criteria were used to produce the most complete sample possible. While previous work has shown that L0 dwarfs have a median color of  $i-z$ = 1.84 \citep{West2008}, we used a bluer color to ensure that any spread of L0 colors was included while mid-M dwarfs were excluded. We did not restrict the $r-i$ color because the $r$-band photometry of these dwarfs is faint and often unreliable. We then used the Hammer spectral typing facility \citep{Covey2007} to analyze each spectrum and assign a spectral type. The Hammer first calculates an initial spectral type based on several molecular and atomic indices, then allows direct comparison of the spectrum to standard templates. We reviewed each spectrum by eye to confirm spectral types and to exclude objects with insufficient SNR for classification, allowing us to construct the most complete sample possible from the SDSS spectroscopic database.

\begin{figure}
\plotone{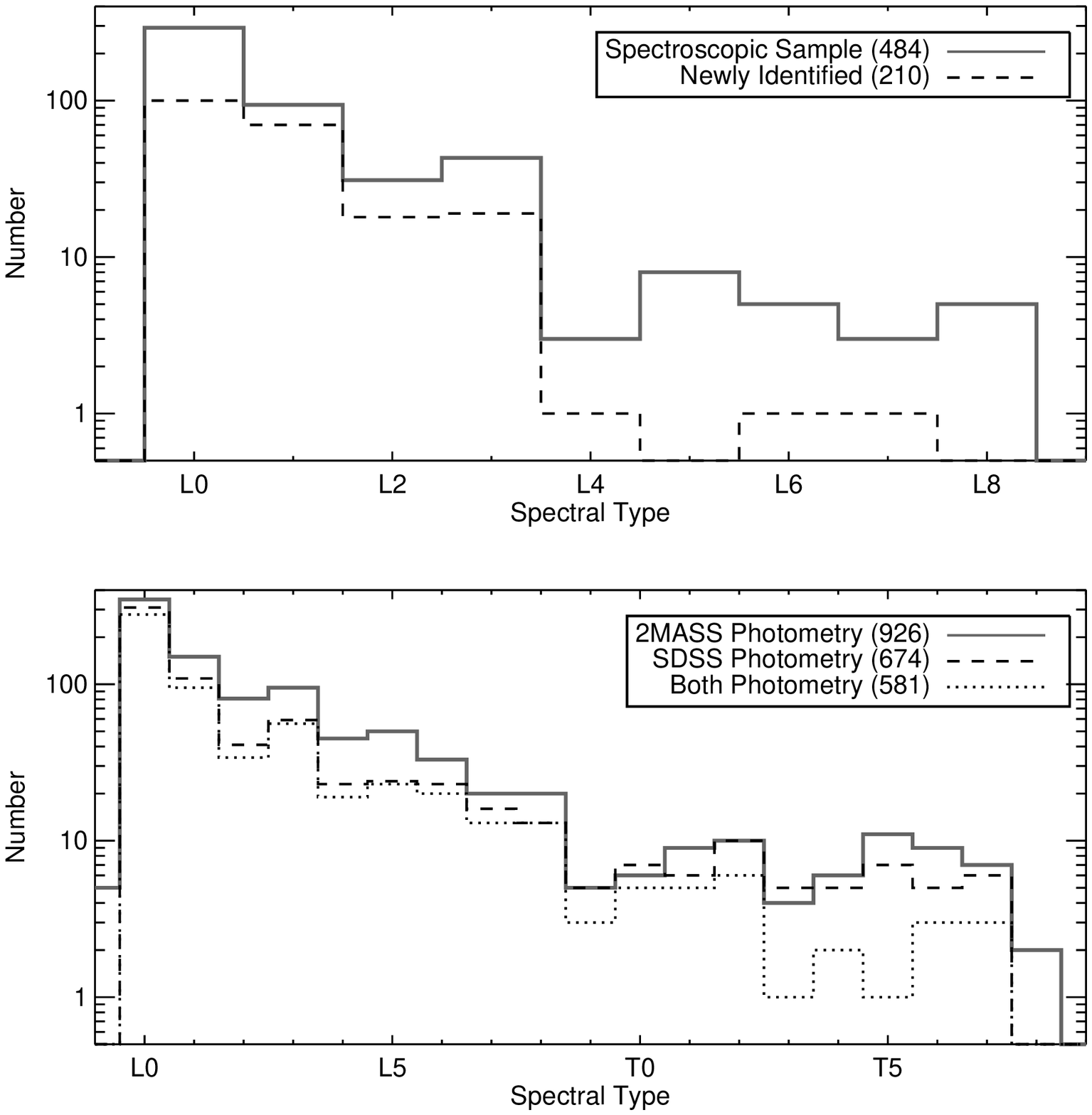} %SDSSL/samhist.pro
\caption{Top: number of objects per spectral type for the spectroscopic sample. The entire sample (grey solid line) and the new dwarfs (black dashed line) are shown. Bottom: number of objects per spectral type bin for the photometric sample. The dwarfs with 2MASS photometry (grey solid line), SDSS photometry (black dashed line), and with both SDSS and 2MASS photometry (black dotted line) are shown.} \label{fig:sthisto}
\end{figure}

Of the initial 13,874 objects, we were unable to classify $\sim$9000 objects due to low SNR, $\sim$4000 were late-M dwarfs, and 484 were L dwarfs. Our selection criteria did not yield any T dwarfs. This sample of 484 L dwarfs represents an order of magnitude increase from H02, which uncovered 42 L dwarfs from SDSS DR1 spectra. We refer to the 484 L dwarfs selected from SDSS spectra as the spectroscopic sample. Their spectral types (typically good to $\pm$1 subtype) are given in Table~\ref{tab:spec} and their spectral type distribution is shown in the top panel of Figure~\ref{fig:sthisto}. The spectroscopic sample is heavily skewed towards early-L dwarfs \citep[compared to the L dwarf luminosity function;][]{Cruz2007} because they are both more numerous and intrinsically brighter than later-L dwarfs. Of the 484 L dwarfs in our spectroscopic sample, 210 are newly discovered, which represents a $\sim$25\% increase in the number of known L dwarfs. Of the previously known dwarfs, agreement with previous spectral types is generally good (within 0-2 types) but $\sim$60 objects that we visually classified as L0 dwarfs were classified by \citet{West2008} as M9 dwarfs. Our visual inspection is more certain than the automated algorithm used in \citet{West2008}, but was not practical given the size of the \citet{West2008} sample. This discrepancy is within the published uncertainties of the Hammer automatic algorithm \citep[$\pm$1 spectral type;][]{Covey2007} 

There are five known L dwarfs with SDSS spectra and that were included in our initial 13,874 objects but were not identified as L dwarfs during spectral typing. Three of them were typed as M9 dwarfs instead of L0 dwarfs so they were excluded from this analysis by our spectral type cut. Another two had SDSS spectra with very low SNR and we were not able to assign a spectral type. These low SNR objects are included in the photometric sample (see Section~\ref{sec:phot}), but not the spectroscopic sample.

\subsection{Activity}
The presence and strength of H$\alpha$ emission is used as an indicator of chromospheric activity in low mass stars and brown dwarfs \citep[e.g.,][]{Gizis2000,West2004,Schmidt2007}. The low SNR of our spectra in the region surrounding H$\alpha$ prevented us from a detailed study of activity in the SDSS L dwarf sample. Out of 484 L dwarfs in the spectroscopic sample, only 32 have SNR $>$ 3 in the H$\alpha$ region. In Table~\ref{tab:act}, we list measured H$\alpha$ equivalent widths (EW) and activity classifications for those dwarfs.

H$\alpha$ EW were measured using trapezoidal integration to calculate the flux in both the line and continuum regions. Following \citet{West2004}, stars that we classified as active must have a measured H$\alpha$ EW greater than 1\AA~and meet three quality control criteria: (1) a measured EW greater than its uncertainty; (2) an H$\alpha$ EW greater than the EW of the comparison regions; and (3) a measured height of the line three times larger than the noise at line center. Objects with H$\alpha$ EW greater than 1\AA~but only meeting some of the quality control criteria were classified as maybe active. Of the 32 L dwarfs, 23 were classified as active, 6 as maybe active, and 3 as inactive. This fraction of active objects ($\sim$70\%) is consistent with results for early-L dwarfs \citep{West2004,Schmidt2007} and reflects a decline from the nearly 100\% active fraction observed in nearby late-M dwarfs \citep{West2008}.

None of our L dwarfs have notably large H$\alpha$ EW, but we report H$\alpha$ detections for some objects previously observed to have emission (noted in Table~\ref{tab:act}). 2MASS J0746425+200032 has been observed in a multi-wavelength campaign by \citet{Berger2009} to have periodically variable H$\alpha$ emission with strength of 2.4-3.3\AA. Our measurement of 2.4$\pm$0.2\AA~is consistent with the previous detections. LHS 2924, a well known active M9 dwarf \citep[e. g.,][]{Fleming1993,Reiners2007a}, was classified as an L0 dwarf during the spectral typing of our sample, which is within our $\pm$1 subtype uncertainty. The H$\alpha$ EW of 4.5$\pm$0.3\AA~measured from its SDSS spectrum is consistent with previous quiescent values.

\subsection{Photometric Sample}
\label{sec:phot}
SDSS $iz$ and 2MASS $JHK_S$ photometry for the spectroscopic sample is given in Table~\ref{tab:spec}. Every object in the sample had SDSS photometry, but we excluded photometry for nine objects with SATURATED, BAD\_COUNTS\_ERROR, INTERP\_CENTER, PSF\_FLUX\_INTERP, or NO\_DEBLEND flags set in the $i$ or $z$ band \citep{Stoughton2002}. Photometry from 2MASS was obtained via a cross-match to the 2MASS point source catalog with a search radius of 5\arcsec; no objects in the spectroscopic sample had more than one match in the 2MASS database. A total of 471 L dwarfs matched 2MASS sources, but we excluded 2MASS photometry for 35 objects with ph\_qual or cc\_flg flags set in the $J$, $H$, or $K_S$ bands \citep{Cutri2003}. 

To perform a more complete analysis of L dwarf colors and examine color trends into the T dwarf regime, we augmented the spectroscopic sample with known L and T dwarfs from the Dwarf Archives\footnote{Available at \url{http://dwarfarchives.org}.} as of October 2009 to construct the photometric sample. For these additional objects, we used previously assigned spectral types (we preferred optical spectral types and rounded half types to earlier whole types) and excluded subdwarfs. Using the same flag cuts as for the spectroscopic sample, we obtained $iz$ magnitudes from SDSS for 148 L and 51 T dwarfs. 2MASS photometry was available for 396 L and 63 T dwarfs, with totals of 48 L and 23 T dwarfs having both SDSS and 2MASS photometry. In four cases ($\sim$1\%), the cross-match between Dwarf Archives and 2MASS returned more than one match. For those objects, we selected the closest source to the input position. We found no obvious color outliers in the sample that would signal a mismatch between objects. The spectral type distribution of photometric sample, including those in our spectroscopic sample, is shown in the bottom panel of Figure~\ref{fig:sthisto}. 

\section{Colors}
\label{sec:color}
\subsection{Color-Spectral Type Relations}
In order to examine the colors of the L and T dwarfs in our photometric sample, we made additional cuts for good quality photometry (photometric uncertainties less than 0.08, 0.06, 0.12, 0.12, and 0.16 for $i$, $z$, $J$, $H$, and $K_S$ respectively) in each band. These uncertainty limits are similar to limiting magnitudes -- the majority of the photometry that is rejected is at or past the limiting magnitudes of SDSS and 2MASS. This cut could preferentially reject objects that are redder in $i-z$ color, because the average SDSS magnitude limits in the $i$- and $z$- band (21.3 and 20.5 respectively) may exclude dwarfs with redder $i-z$ colors. Comprehensive follow up of the redder sources is needed to confirm median colors, especially for the later-L dwarfs.

\begin{figure}
\includegraphics[width=0.75\linewidth]{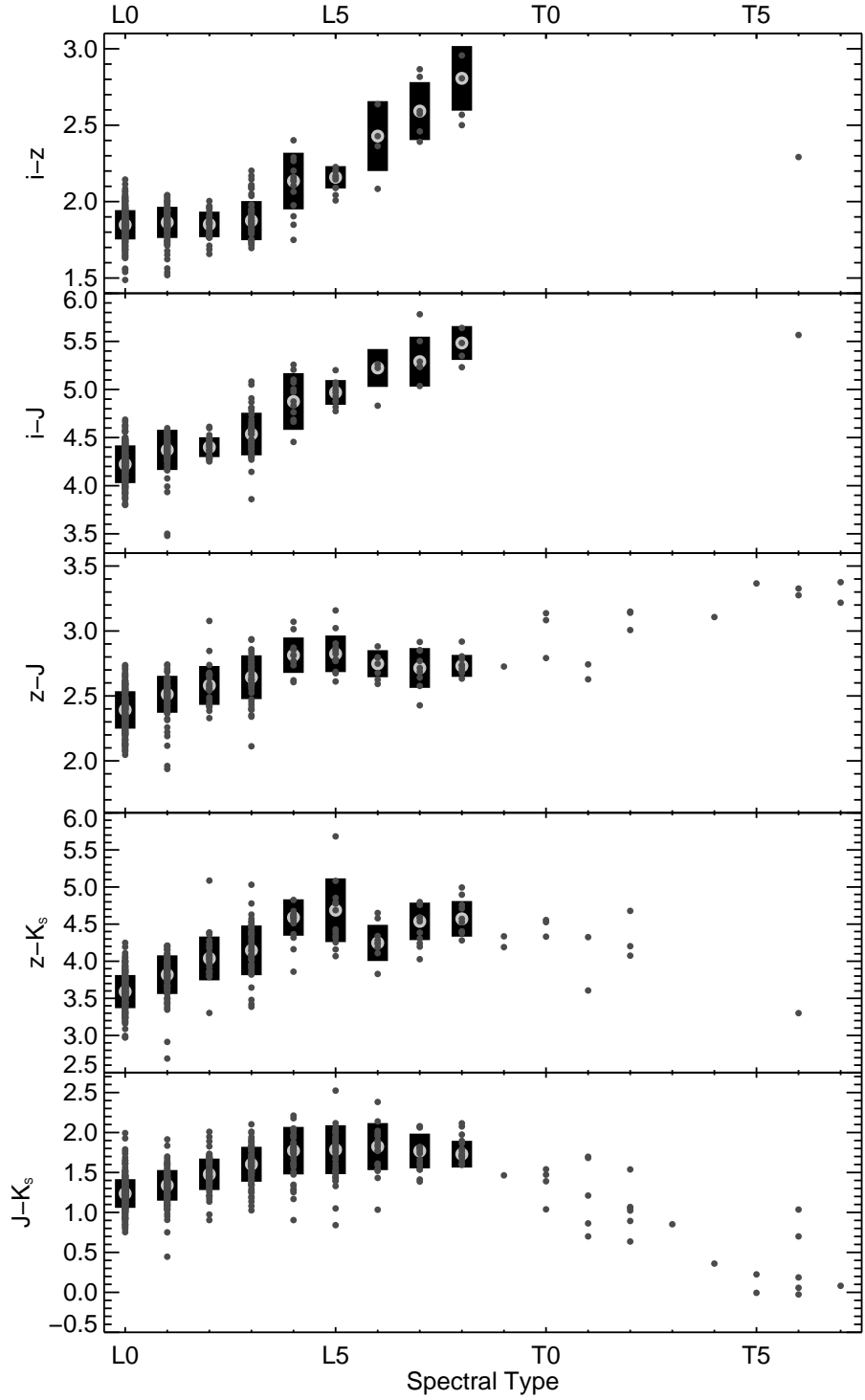} %SDSSL/colors/gtab.pro
\caption{Color as a function of spectral type for five different colors. Individual dwarfs are shown (small dark grey circles) for L0-T7 dwarfs, as well as median colors (large light grey circles) and one sigma dispersions (black bars) for L0-L8 dwarfs.} \label{fig:stcol}
\end{figure}

Examining the relationship of broadband colors to spectral type is useful both to provide insight into mean L dwarf properties and to assist future searches for L dwarfs. Table~\ref{tab:col} lists median colors as a function of spectral type for a variety of SDSS and 2MASS colors. For each spectral type, we give the number of objects in the photometric sample with good photometry that contribute to each color, the median color with associated uncertainty, and the standard deviation ($\sigma$). The uncertainties in the median were calculated using the standard deviation divided by the the square root of the total number of stars in the sample. While the uncertainty reflects how well we know the median color of each spectral type bin, the standard deviation reflects the intrinsic scatter in each color. Figure~\ref{fig:stcol} shows color as a function of spectral type for the same SDSS and 2MASS colors. The median colors with standard deviation are shown for types L0-L8. 

As has been noted by previous studies \citep[H02;][]{Knapp2004,Chiu2006}, the $i-z$ color gets redder at increasing spectral type -- rising slowly from L0-L3 and then rapidly with later spectral type through mid-T dwarfs. This is because the $i$-band is centered near the K I doublet (7665, 7699\AA), which broadens with decreasing temperature and suppresses proportionally more flux in the $i$-band than the $z$-band. The $i-J$ color also becomes redder with later spectral type, but changes more linearly than $i-z$ \citep[H02;][]{Sheppard2009}. Both $i-z$ and $i-J$ are relatively good predictors of spectral type.

The $z-J$ color becomes redder from early- to mid-L, stays constant from mid- to late-L types, and reddens slightly from late-L to late-T. This is consistent with both the SDSS/MKO $z-J$ \citep{Knapp2004,Chiu2006} colors and SDSS/2MASS $z-J$ colors than found in H02. Because of this behavior, $z-J$ color is a less reliable predictor of spectral type than $i-z$ and $i-J$ throughout the L spectral sequence.

Both the $z-K_S$ and $J-K_S$ color redden until mid-L types, then turn over and become more blue into the late-T \citep[H02;][]{Knapp2004,Chiu2006,Faherty2009}. This behavior is likely due to the decrease in $K_S$-band flux due to increasingly strong absorption in water and methane bands \citep{Marley2002,Knapp2004}.

\subsection{Red Bias in $J-K_S$ in Previous Samples} 
\label{sec:redb}
\begin{figure}
\plotone{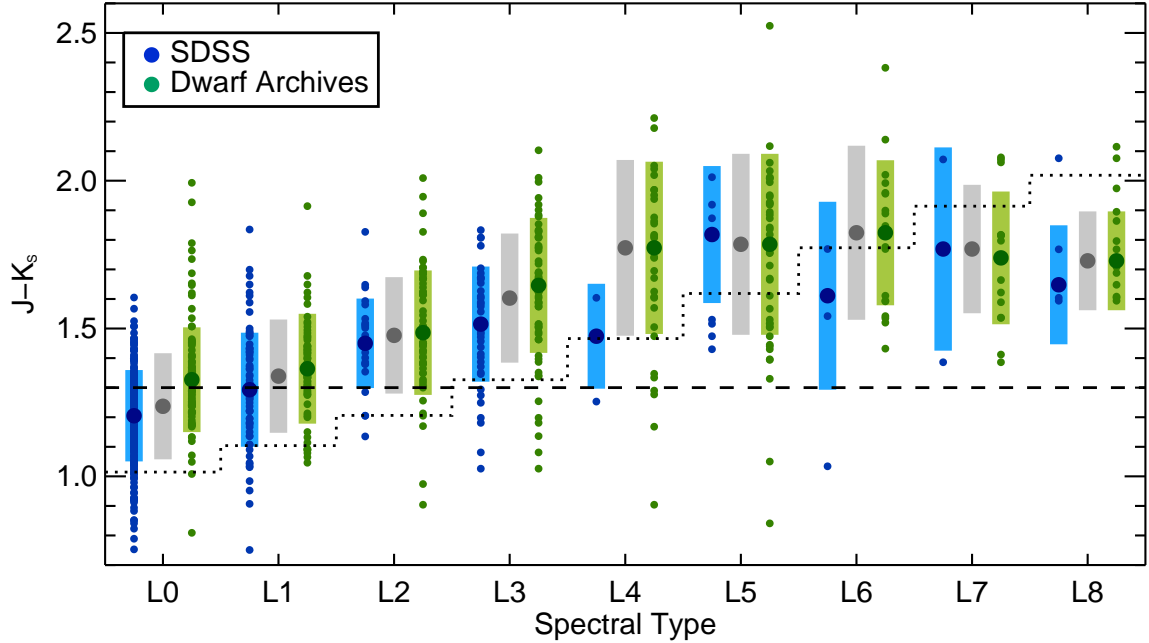} %SDSSL/colors/mkmedtab.pro
\caption{$J-K_S$ color as a function of spectral type for the SDSS dwarfs (blue), all dwarfs (grey) and dwarfs from Dwarf Archives as of October 2009 (green). The median color per spectral type for each sample is shown (large circles) as well as the standard deviation of the colors in that spectral type (shaded bars) and the individual objects (small circles). The \textit{J$-$K$_S$} color selection criteria of  \citet[dashed line]{Kirkpatrick1999} and \citet[dotted line]{Cruz2003} are shown; the latter is a combination of a color-magnitude cut and the \citet{Cruz2003} spectral type vs.\ M$_J$ relation for dwarfs at 20~pc.} \label{fig:redb}
\end{figure}

The median colors for L0 dwarfs in the spectroscopic sample are consistent with those given by \citet{West2008}, and the SDSS/2MASS colors for the rest of the dwarfs are similar to those in previous work \citep[H02;][]{Knapp2004,Chiu2006}, but the $J-K_S$ colors calculated from the photometric sample are consistently bluer than those found in previous work \citep{Kirkpatrick2000,Faherty2009}. 
Figure~\ref{fig:redb} shows the $J-K_S$ colors of all L dwarfs with 2MASS photometry compared to both the L dwarfs in the spectroscopic sample and the L dwarfs in Dwarf Archives; the data are given in Table~\ref{tab:JK}. Early-L dwarfs in the SDSS spectroscopic sample are clearly bluer than the Dwarf Archives objects by $\sim$0.1 magnitude.

While the spectroscopic sample is likely to have no bias in $J-K_S$ color, the objects from Dwarf Archives were selected from a variety of criteria, including large numbers found using 2MASS color criteria. Many known L dwarfs are the result of surveys that made cuts in $J-K_S$ to separate L dwarfs from the ubiquitous M dwarfs \citep{Kirkpatrick1999,Kirkpatrick2000,Cruz2003,Cruz2007,Reid2008}. The color selection criteria from \citet{Kirkpatrick1999} and \citet{Cruz2003} are shown in Figure~\ref{fig:redb}. The \citet{Kirkpatrick1999} cut at $J-K_S > 1.3$ excludes many early-L dwarfs, and some objects at mid-L types. The \citet{Cruz2003} cut, though slightly bluer than the median color, still excludes some of the bluest objects at each type.

To test whether previously known L dwarfs are drawn from a different distribution than the SDSS spectroscopic sample, we performed a Kolmogorov-Smirnov test (KS test) on the color distribution of the two samples. In order to examine all of the L dwarfs as a single population, we used a color difference that is independent of spectral type, defined as  $\delta_{J-K_S} = ((J-K_S) - (J-K_S)_{med})/\sigma_{J-K_S}$. We used the median $J-K_S$ and $\sigma_{(J-K_S)}$ for all L dwarfs given in Table~\ref{tab:col}. The KS test shows that the two samples are \emph{not} drawn from the same distribution at a high confidence level (P$<$0.01).

This selection bias can lead to systematic problems when $J-K_S$ color criteria are used to assemble large statistical samples of L dwarfs. There is a strong potential to miss nearby blue objects \citep[Section~\ref{sec:new};][]{Schmidt2010,Bowler2010} and to bias all known L dwarfs towards redder colors. With mounting evidence that age and $J-K_S$ color are linked \citep[Section~\ref{sec:jkout};][]{Folkes2007,Burgasser2008a,Faherty2009}, previous samples are likely biased towards younger dwarfs. Past kinematic studies have noted that the L dwarf population seems unusually young \citep[e.g., ][]{ZapateroOsorio2007} -- this could be in part due to a color selection bias. 

\subsection{Color-color relations}
\begin{figure}
\plotone{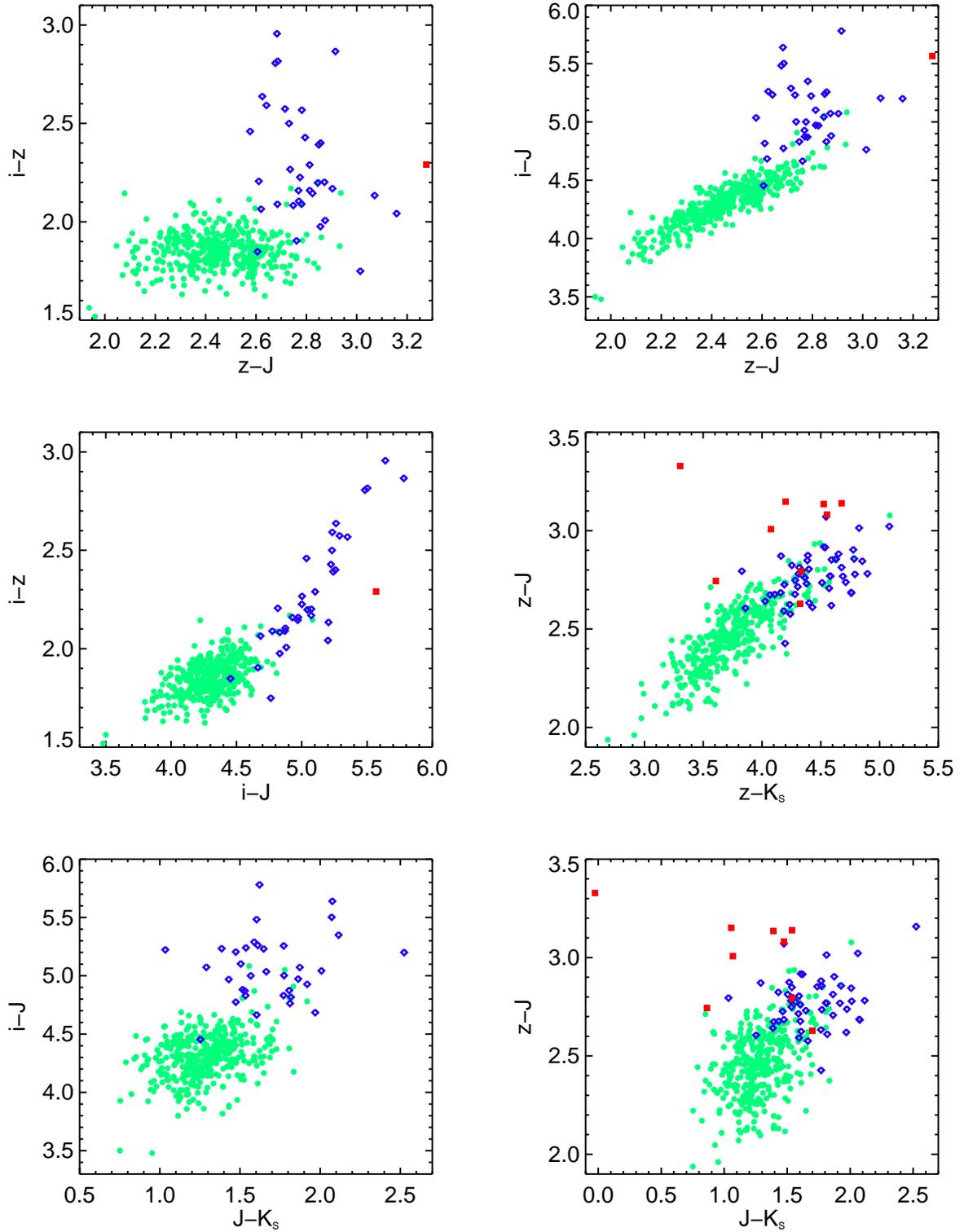} %SDSSL/colors/gcol.pro
\caption{Color color plots for six different color combinations. Early-L dwarfs (L0-L3; green circles), mid- to late- L dwarfs (L4-L8; blue diamonds) and T dwarfs (red squares) are distinguished.} \label{fig:color}
\end{figure}

Figure~\ref{fig:color} shows six color-color plots for our combined photometric sample. To examine the different color spaces for each spectral type, we distinguish between early-L, late-L, and T dwarfs. 

The first four panels show updated color-color plots similar to those from the SDSS Early Data Release (H02). The $i-z$, $z-J$ and $i-J$, $z-J$ plots show the same general shape as in H02. The $i-z$ and $i-J$ colors are relatively linear with spectral type. The $z-J$, $i-J$ plot is useful for classifying L0 to L3 dwarfs, but the $z-J$ color saturates for mid- to late- L dwarfs. The $i-z$, $i-J$ plot shows a remarkably straight line in color-color space for $i-J > 4.8$ and is most useful for classifying L4 and later types. In $z-J$,  $z-K_S$ the early-L dwarfs are spread out, while the late-L and T dwarfs are clumped. A turnover in $z-K_S$ color is also apparent, similar to the one seen in H02.

The last two panels show $i-J$ and $z-J$ as a function of $J-K_S$. Both $i-J$ and $z-J$ colors separate the early-L and T dwarfs into distinct groups despite their similar $J-K_S$ colors, and can be used to roughly classify objects as early-L, late-L or T dwarfs. This shows the value of including a bluer photometric band together with 2MASS colors when analyzing L and T dwarf photometry.

\subsection{Photometric Distance Estimates}
\label{sec:phtplx}
While spectral type is a relatively reliable predictor of absolute magnitude for L dwarfs \citep[e.g.,][]{Vrba2004}, spectrophotometric absolute magnitudes and distance estimates suffer from the uncertainty and coarse binning of spectral types, which are often good only to $\pm$1 subtype. When only 2MASS colors are available, spectral type relations are better because of the large spread in infrared colors with both absolute magnitude and spectral type. However, the $i-z$ and $i-J$ colors show much smaller scatter than the spectral type relations.

\begin{figure}
\plotone{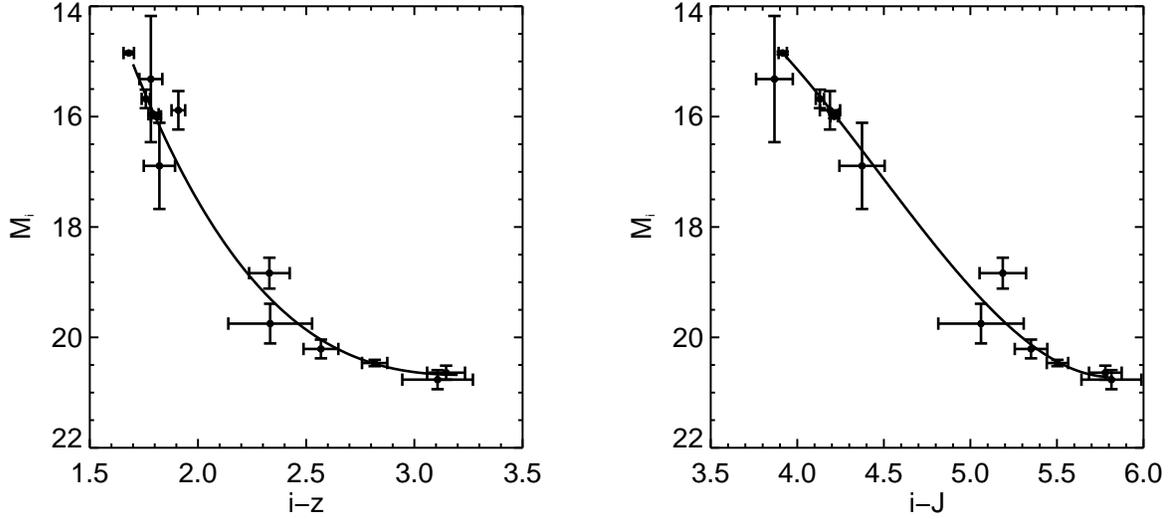} %SDSSL/kine/phtplx.pro
\caption{Absolute $i$ magnitude as a function of $i-z$ (left) and $i-J$ (right) color, for M8-L8 dwarfs with SDSS photometry (given in Table~\ref{tab:plx}). Third-order polynomial fits are shown (equations given in Section~\ref{sec:phtplx}).} \label{fig:fit}
\end{figure}

Figure~\ref{fig:fit} illustrates M$_i$ as a function of $i-z$ and $i-J$ color based on photometry and parallax measurements for the 13 late-M and L dwarfs listed in Table~\ref{tab:plx}. The reduced $\chi^2$ value for each of the relations was minimized with third order polynomial fits, as follows:
\begin{equation}  (1.7 <i-z<3.2): M_i = -23.27 + 38.41(i-z) - 11.11(i-z)^2  + 1.064(i-z)^3 \end{equation}
\begin{equation} (3.9 < i-J < 5.8): M_i = 66.88 - 41.73(i-J) + 10.26(i-J)^2 - 0.7645(i-J)^3 \end{equation}

\noindent The relations have RMS scatters of 0.41 and 0.33 magnitudes in $i-z$ and $i-J$ respectively.

\begin{figure}
\plotone{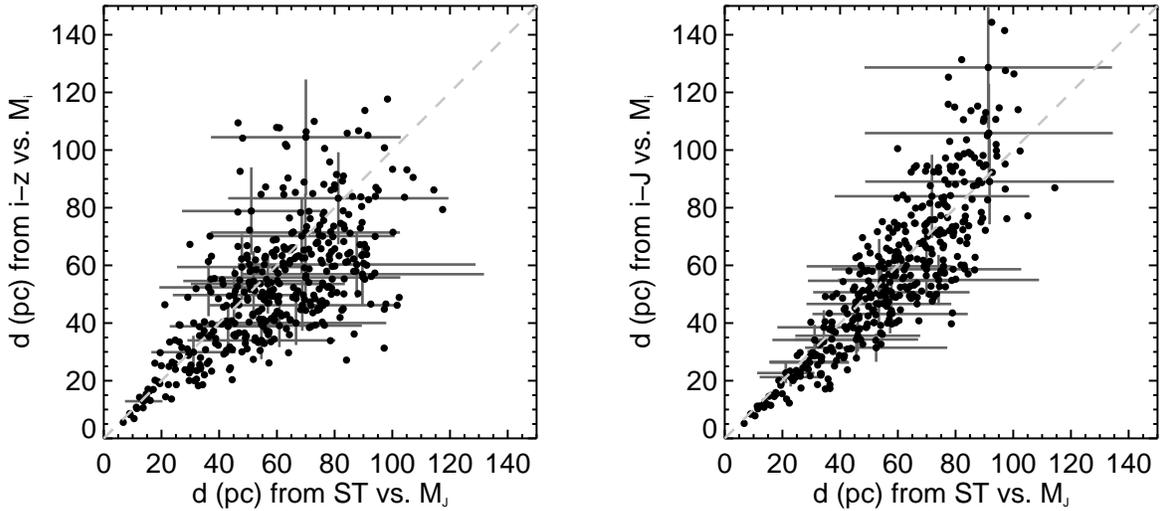} %SDSSL/kine/dcomp.pro
\caption{Comparison of spectrophotometric distance estimates using the \citet{Cruz2003} ST vs. M$_J$ relation with distance estimates from our $i-z$ vs. M$_i$ (left panel) and  $i-J$ vs. M$_i$ (right panel) relations. Uncertainties for a representative number ($\sim$5\%) of the dwarfs (grey bars) and a one-to-one correspondence line are shown (grey dashed).} \label{fig:dcomp}
\end{figure}

Distances computed from the derived absolute magnitudes are given in Table~\ref{tab:kine}. We compare distance estimates from both of these relations to those from the \citet{Cruz2003} spectral type vs.\ M$_J$ relation in Figure~\ref{fig:dcomp}. There is a wide scatter about the one-to-one line, analogous to the scatter between $i-z$ or $i-J$ and spectral type, but 95\% of the distance estimates agree within their uncertainties. Distances above 100~pc calculated from the $i-J$ vs.\ M$_i$ relation seem to show a systematic offset from distances calculated from the spectral type vs.\ M$_J$ relation. The dwarfs with these disparate distances estimates all have $i-J$ colors on the bluer end of the relation (3.9 $< i-J <$ 4.2). The $i-J$ vs.\ M$_i$ relation is steep in that color range, and while those $i-J$ colors are within the spread for L0/L1 dwarfs, the relation is also based on colors from M8/M9 dwarfs. This could be the reason for the disagreement between the distance estimates.

For the remainder of the paper, we use the $i-z$ distance estimates due to the better agreement with previous distances. For L dwarfs that fall outside the $i-z$ color limits given in Equation 1, we use the $i-J$ relation. For dwarfs outside both color limits but with L spectral types, we use the spectral type vs.\ $M_J$ relation from \citet{Cruz2003}.

\subsection{New L Dwarfs Within 30~pc}
\label{sec:new}
Previous searches for L dwarfs have focused on a census of nearby L dwarfs, probing distances of 20-30~pc from the Sun. Due to the comprehensive work done at these distances, the majority of newly discovered L dwarfs presented in this paper are early-L dwarfs at distances greater than 30~pc. There are, however, a few nearby dwarfs that were missed by previous selection criteria. Table~\ref{tab:close} shows data for 5 L dwarfs which are placed within 30~pc of the Sun by the mean of all three distance estimates, weighted by the distance uncertainties. 

Of the five L dwarfs, three are unusually blue for their spectral types, which is unsurprising given the bias of previous searches towards dwarfs with redder $J-K_S$ colors (Section~\ref{sec:redb}). The nearest and bluest of these is SDSS J141624.09+134826.7 (hereafter SDSS1416+13), an exceptionally blue L dwarf within 10~pc of the Sun. Additional spectroscopic observations \citep{Schmidt2010} show that it is an L5 dwarf with an updated distance estimate of 8.9$\pm$2.1~pc. 

The discovery of these new objects at $d<30$~pc indicates that there is potential to discover additional nearby L dwarfs. While SDSS is an excellent tool for uncovering dwarfs with peculiar 2MASS colors, the SDSS footprint does not cover the entire sky, and as yet there has been no comprehensive spectroscopic follow-up of all objects with L dwarf colors. 

\section{Kinematics}
\label{sec:kine}
\subsection{Proper Motions}
Our combined photometric sample contains 586 objects that have both SDSS and 2MASS photometry. These two surveys are astrometrically calibrated and span a range of dates (2MASS 1997-2001; SDSS 2000-2008), which allowed us to calculate reliable proper motions for the majority of objects in our sample, using the difference between the SDSS and 2MASS positions. Objects were cross-matched using a search of the SDSS coordinates in the 2MASS database with a radius of 5\arcsec. None of the objects with SDSS coordinates had multiple matches in the 2MASS database, and we found no mismatches by looking at outliers in color space. The number of mismatches between the two databases is likely low.

\begin{figure}
\includegraphics[width=0.8\linewidth]{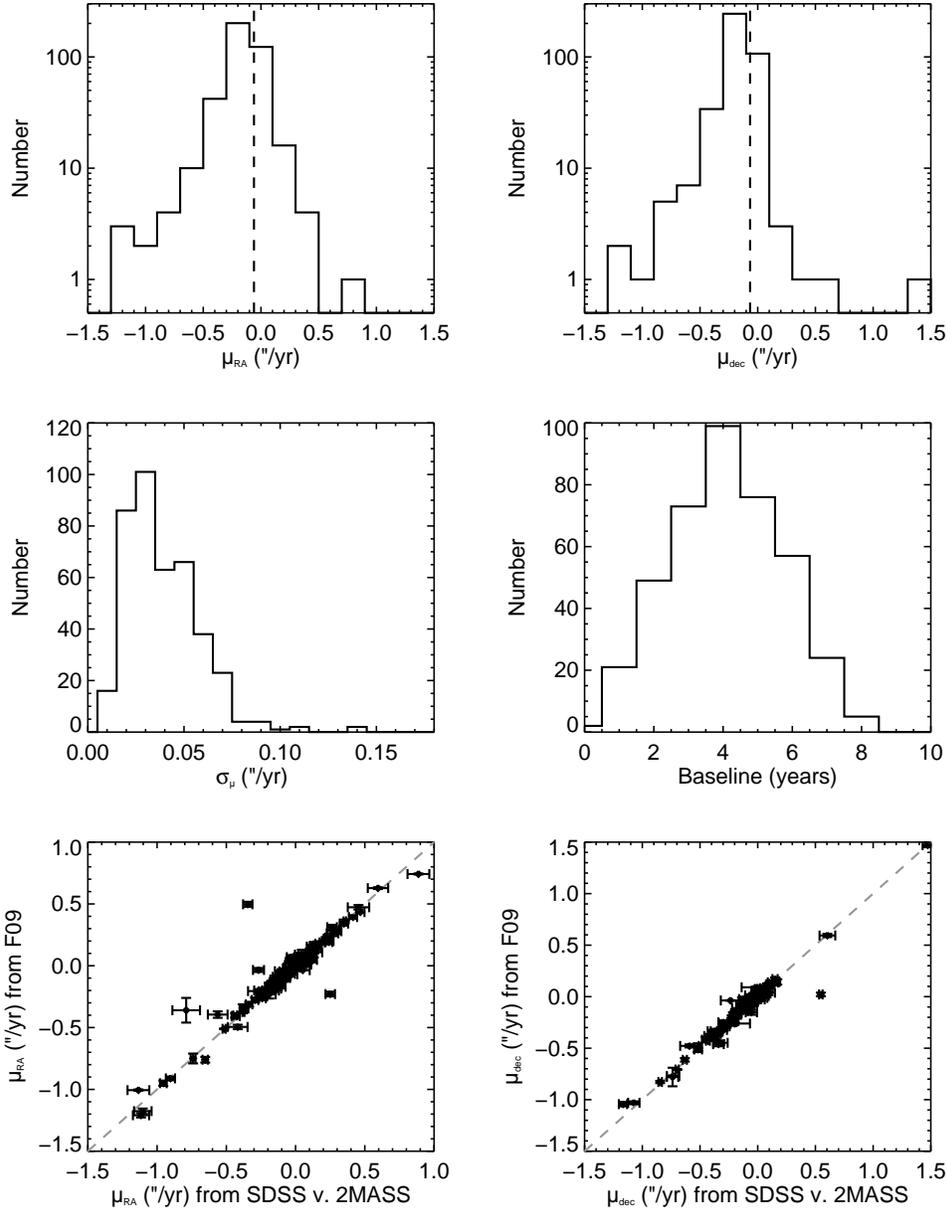} %SDSSL/kine/pmplot.pro
\caption{Top panels: Histrograms showing the distribution of measured proper motions in RA (top left) and dec (top right). In both panels, the mean proper motion is shown (dashed line). Middle left panel: The distribution of total proper motion uncertainties.  Note that cuts have been made of $\sigma_{\mu} < 0.8$ "/yr or $\sigma_{\mu}/\mu_{tot} < 0.25$. Middle right panel: The distribution of baselines used to measure proper motions. Bottom two panels: Comparison of our proper motions in RA (left) and dec (right) to measurements from Faherty et al.\ (2009)
and references therein. A one-to-one correspondence line is shown.} \label{fig:pms}
\end{figure}

Proper motions were calculated for all objects with two sets of coordinates, but we only include measurements that had total proper motion uncertainties ($\sigma_{\mu}^{1/2} = (\sigma_{RA}^2 + \sigma_{dec}^2)^{1/2}$) of less than $\sigma_{\mu} < 0.08"/yr$ or $\sigma_{tot}/\mu_{tot} < 0.25$. The inclusion of dwarfs with $\sigma_{tot}/\mu_{tot} < 0.25$ allows for objects with relatively large proper motions to be included if their uncertainties are small compared to their proper motions. This should not bias the sample heavily towards faster moving objects because it only includes 14 dwarfs that would have been otherwise excluded. Table~\ref{tab:kine} gives proper motions for the 406 dwarfs (312 in the spectroscopic sample) that have reliable calculated proper motions. The distributions of $\mu_{\alpha}$, $\mu_{dec}$, $\sigma_{\mu}$, and time between observations are shown in Figure~\ref{fig:pms}. 

Of the 406 objects with measured proper motions, 135 have previous measurements \citep[and references therein]{Faherty2009}. The bottom two panels of Figure~\ref{fig:pms} show the comparison of our measured proper motions to those from literature. There are only 9 objects that disagree by more than 0.08\arcsec/yr in total proper motion. We were unable to calculate SDSS/2MASS proper motions for 35 objects in the spectroscopic sample that have measurements available from the literature. We include these literature measurements (which are noted in Table~\ref{tab:kine}) in order to calculate three dimensional velocities for these objects.

\subsection{Tangential Velocities}
\begin{figure}
\plotone{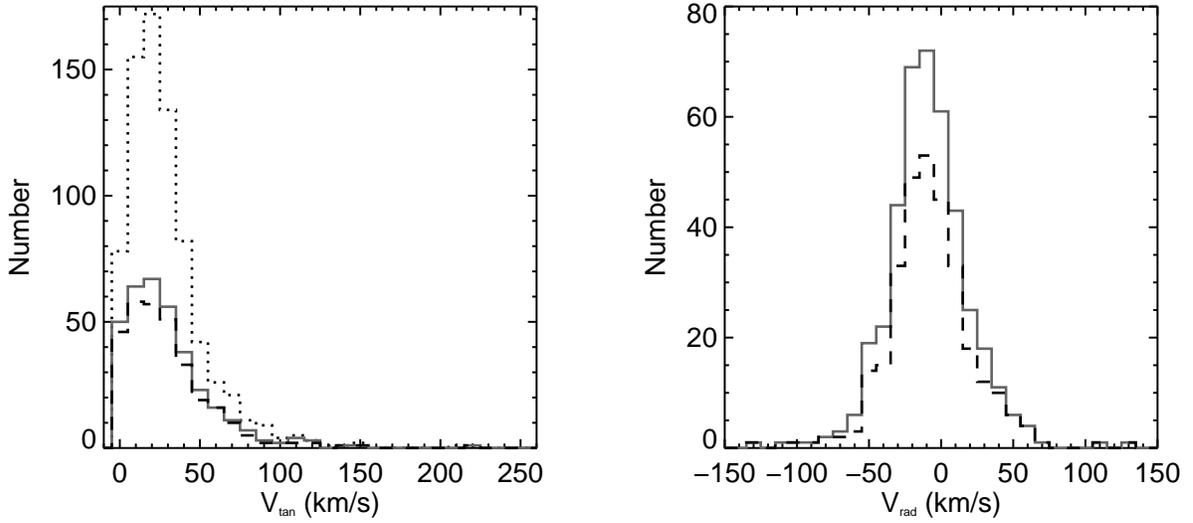} %SDSSL/kine/vhisto.pro
\caption{Histogram of tangential (left) and radial (right) velocites. In both panels, all dwarfs with complete $UVW$ motions are shown (dashed lines). Dwarfs in the spectroscopic sample that only have tangential velocity measurements (left) or radial velocity measurements (right) are shown as the solid line. In the left panel, all dwarfs (including those previously known) with proper motions and distance estimates are also shown as the dotted line.} \label{fig:vtan}
\end{figure}

While three dimensional velocities provide more accurate kinematic results, we could only measure reliable radial velocities for 413 dwarfs in the spectroscopic sample (see Section~\ref{sec:radv}) and have no radial velocities for our photometric sample. By combining proper motions and distances from this paper and from the literature, we examined the tangential velocity distribution of 748 L dwarfs, shown in the left panel of Figure~\ref{fig:vtan}. The entire sample of velocities has a median of V$_{\rm tan}$ = 28~km~s$^{-1}$ and a dispersion of $\sigma_{\rm tan}$ = 25~km~s$^{-1}$. These values are consistent with previous results for L dwarfs \citep{Faherty2009,Schmidt2007}.

The tangential velocity distributions for 347 L dwarfs in the spectroscopic sample and for 306 dwarfs with complete $UVW$ kinematics are also shown in the left panel of Figure~\ref{fig:vtan}. They have dispersions of $\sigma_{\rm tan}$ = 28~km~s$^{-1}$ and $\sigma_{\rm tan}$ = 27 ~km~s$^{-1}$ respectively, which are consistent both with previous work and with the larger sample of L dwarfs discussed above.

\subsection{Radial Velocities}
\label{sec:radv}
Ten L dwarfs with spectra in the SDSS database also have radial velocities measured with high resolution spectroscopy  \citep{Mohanty2003,BailerJones2004,Blake2007}. We used these SDSS spectra as radial velocity templates. The template spectra span the entire L-dwarf spectral sequence with gaps no larger than one spectral type. We cross-correlated each spectrum from the spectroscopic sample with every template within one spectral type (in some cases there was only one such template). The minimum of the cross-correlation function was fit to determine the radial velocity.  Each cross-correlation function was examined by eye to determine its quality.  Cross-correlations that did not yield acceptable minima were skipped.  All others were ranked according to the depth and symmetry of their form.  Symmetric cross-correlation functions with troughs lower than 50\% of the wings were assigned double the weight when computing the mean radial velocity for each spectrum.  Weighted standard deviations were also determined for each spectrum with more than one cross-correlation template within one spectral type. Measurements with uncertainties greater than 20~km~s$^{-1}$ were not used. Dwarfs cross-correlated with only one spectrum had no formal uncertainties, so we conservatively assign them uncertainties of 20~km~s$^{-1}$.

Radial velocities are given in Table~\ref{tab:kine} and the distribution of radial velocities is shown in the right panel of Figure~\ref{fig:vtan}. The radial velocity distribution has a mean of $<V_{\rm rad}>=-4.7$~km~s$^{-1}$ and a dispersion of $<\sigma_{\rm rad}>=34.3$~km~s$^{-1}$. The deviation of the mean of the radial velocity distribution from $<V_{\rm rad}>=0$~km~s$^{-1}$ is likely the effect of Solar motion, and not due to a systematic error in our measurements.  Most of the sight lines to our L dwarfs are toward high northern Galactic latitudes (where most of the SDSS sky coverage is).  In addition, the Sun's motion out of the Plane is $W$ = 7 km s$^{-1}$ \citep{Dehnen1998}.  Therefore, because the mean vertical motion of stars in the Galaxy is near 0 km s$^{-1}$, the distribution of stars in our sample appears to be moving toward us at the Solar vertical motion.  We have corrected for the Solar motion in the W velocities described below.

\subsection{$UVW$ Motions}
Using proper motions, photometric distances, and radial velocities for 306 dwarfs in the spectroscopic sample, we calculated \textit{UVW} velocities and uncertainties with the method outlined in \citet{Johnson1987}. Our velocities (given in Table~\ref{tab:kine}) were corrected to the Local Standard of Rest assuming $UVW_{\odot}$ = (-10, 5, 7) km s$^{-1}$ \citep[with positive $U$ defined toward the Galactic center]{Dehnen1998}. Because our L dwarfs are located within 120~pc, these $UVW$ velocities are essentially equal to their Galactic $V_r$, $V_{\phi}$, and $V_z$ velocities. 

\begin{figure}
\plotone{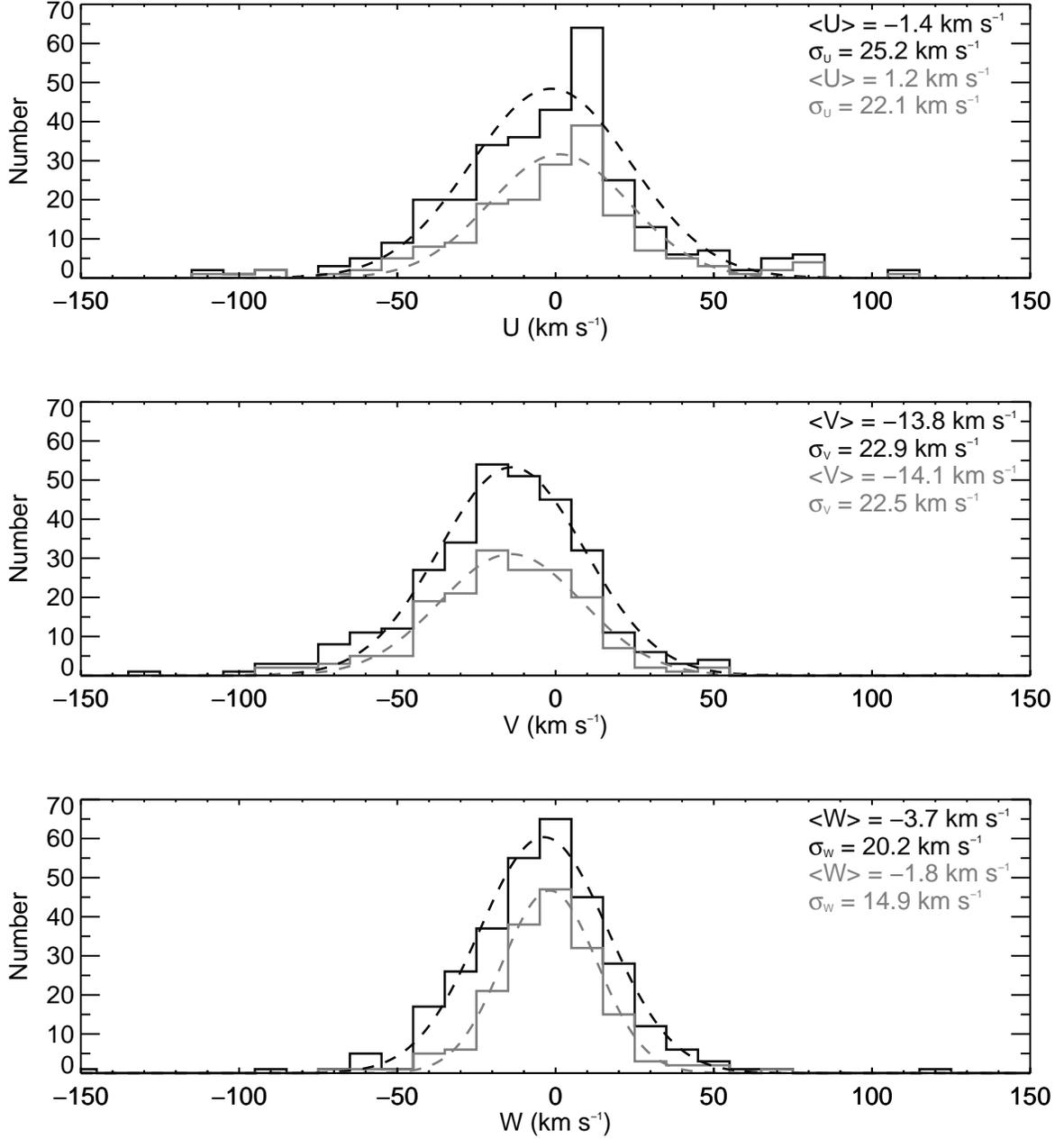} %SDSSL/kine/UVWhist.pro
\caption{$UVW$ histograms for all L dwarfs (black) and for dwarfs with good kinematics (grey). A Gaussian fit is shown (dashed line) for each distribution, and the best fit mean and dispersion for the Gaussian is given in the upper right corner of each plot.} \label{fig:uvw}
\end{figure}

To investigate possible biases due to the inclusion of fast moving objects (because of the proper motion cut, $\sigma_{tot}/\mu_{tot} < 0.25$) and poorer quality measurements, we compared our sample to a subsample of 175 objects with lower uncertainty cuts ($\sigma_{\mu} < 0.08 \arcsec$~yr$^{-1}$ and $\sigma_{rad} < 12$~km~s$^{-1}$). Figure~\ref{fig:uvw} shows $UVW$ histograms of both the total sample and the subsample. We fit Gaussian functions to each of the velocity distributions to determine the mean and standard deviations (shown on each panel of Figure~\ref{fig:uvw}). The total sample shows good agreement with the lower uncertainty subsample; we use the total sample in the remainder of our analysis to take advantage of the larger number of velocities.

If we treat our kinematic distributions as a single population, our dispersions ($\sigma_U,\sigma_V,\sigma_W$) = (25, 23, 20) km~s$^{-1}$ show reasonable agreement with those of previous studies. \citet{ZapateroOsorio2007} found dispersions of ($\sigma_U,\sigma_V,\sigma_W$) = (30, 17, 16) km~s$^{-1}$ from complete kinematics of a sample of 20 L and T dwarfs, and \citet{Faherty2009} calculated ($\sigma_U,\sigma_V,\sigma_W$) = (28, 23, 15) km~s$^{-1}$ from tangential velocities of a distance-limited sample of 114 L dwarfs.

\subsection{Kinematic Models}
Kinematics are useful tools to investigate the ages of populations of stars and brown dwarfs. As a population ages, its members have an increasing number of gravitational interactions with molecular clouds in the plane of the Galaxy. These interactions change the velocity of each star in a random direction and by a random amount, which increases the velocity dispersion of the population. Kinematically cooler populations (smaller dispersions) are typically younger than hotter populations (larger dispersions). The kinematics of disk stars are often fit with two components - a cooler, younger population and a hotter, older population \citep[e.g.,][]{Reid2002}. 

While fitting the L dwarf velocities with a single Gaussian is useful for comparing our population to previous work, a two component analysis is important to examine how the L dwarf population compares to other disk dwarf populations \citep[e.g. M dwarfs;][]{Bochanski2007b}.  With large samples of objects, this can be easily accomplished by fitting two Gaussian functions to a histogram of the velocity distribution.  The cold kinematic population dominates near the mean velocity, while the hotter kinematic population is usually only apparent a few standard deviations from the mean.  For small samples like ours (N $<$ 400), the hotter population is difficult to fit, with the Gaussian fit depending strongly on a few outliers in the wings of the velocity distribution. 

\begin{figure}
\plotone{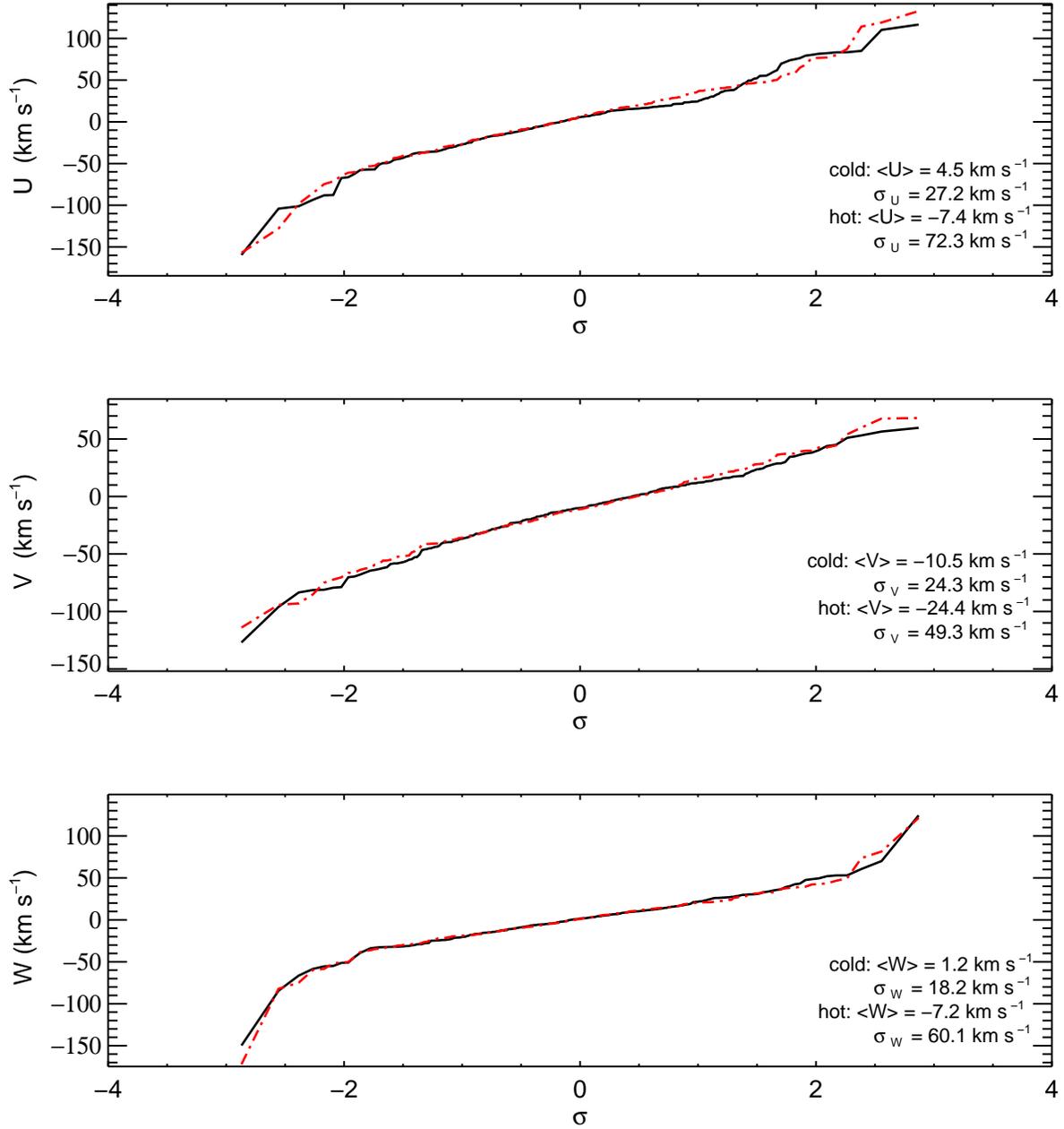} %Sebastian
\caption{Probability plots for $UVW$ data (black) compared to modeled data (red dot dashed). A population fit by a single Gaussian would appear to be a straight line. The means and dispersion for the dynamically hot (10\% of total) and cold (90\% of total) populations are given.} \label{fig:prob}
\end{figure}

This has been addressed in the past by fitting lines to a cumulative probability plot (shown for our data in Figure~\ref{fig:prob}). In a cumulative probability plot, the data are shown as a function of the expected deviation from the mean for a Gaussian distribution; a population well-described by a single Gaussian function would be a straight line \citep{Lutz1980}. The dispersion of the cold population is then given by the slope of a line fit to the data near the mean, and the dispersion of the hot population is given by the average of the slopes of two lines fit to the data a few sigma in either direction from the mean \citep{Reid2002,Bochanski2007b}. While it is evident from Figure~\ref{fig:prob} that our sample contains both a hot and cold kinematic component, there are too few outlying points to use the slope of the wings to obtain velocity dispersions. We have therefore compared the velocity distributions to model kinematic populations.

We generated models by assigning a fraction of the total number of stars to each of the two components. We then randomly populated a comparison data set based on a mean and dispersion for each of the components, using the Box-Muller method \citep{Box1958}.
We used the Levenberg-Marquardt method for least squares fitting \citep{Levenberg1944,Marquardt1963} to minimize the difference between the model distribution and the data by varying the means and dispersions used to generate the model distribution. We repeated this process 1000 times for each fraction. The best fit parameters selected for the fraction are the mean of all the results. The uncertainties in the parameters are the standard deviation of all the results divided by the square root of the total number of stars.
Once the best fit parameters for each fraction were found, we then determined the best fit fraction by choosing the fraction with the minimum average $\chi^2$ fit. Only the $W$ velocity showed a strong minimum $\chi^2$, so we chose that fraction (90\% cold, 10\% hot) for the $UV$ velocities as well.

Comparisons of our model and observed kinematic distributions are shown as cumulative probability plots in Figure~\ref{fig:prob}. The means and dispersions of each of the populations are shown on each panel of Figure~\ref{fig:prob} and given in Table~\ref{tab:2kin}. As expected, the velocity dispersions for the cold component are similar to those obtained by fitting the entire population with one distribution. The dispersions of the cold and hot components are in good agreement with M dwarfs at Galactic heights  $|z| < 100$~pc \citep[in prep.]{Bochanski2007b,Pineda2010} indicating that the L dwarfs are kinematically similar to the local disk M dwarf population. This indicates that the L dwarfs have similar ages to local disk stars, instead of being a kinematically younger population, as has been previously suggested \citep{ZapateroOsorio2007}.

\subsection{$J-K_S$ Outliers and Age}
\label{sec:jkout}
The spread in $J-K_S$ color within each spectral type is much wider than the spread in $i-z$, $i-J$, or $z-J$. Follow-up spectroscopic observations have shown that dwarfs with unusual $J-K_S$ colors are also sometimes spectroscopically peculiar. Red $J-K_S$ outliers often exhibit either distinctive low gravity features, indicating that these objects are young and still collapsing \citep{Kirkpatrick2008,Cruz2009} or evidence of dusty photospheres, which can also be attributed to young ages \citep{Looper2008a}. Blue outliers have been found to show strong H$_2$O and weak CO in their infrared spectra, possibly associated with high gravity and old age \citep{Burgasser2008a}. 

Additionally, a majority of the spectroscopically young red outliers have positions coincident with young moving groups \citep{Cruz2009}, while many of the spectroscopically peculiar blue outliers have fast velocities consistent with being members of an older population. \citet{Faherty2009} investigated the kinematics of two groups of red and blue outliers (regardless of spectroscopic peculiarities), and found that the red outliers (more than 2$\sigma$ from mean $J-K_S$ color) have a smaller tangential velocity dispersion ($\sigma_{\rm tan}$ = 16 km/s) and the blue outliers have a larger tangential velocity dispersion ($\sigma_{\rm tan}$ = 47 km/s) compared to the full sample dispersion of $\sigma_{\rm tan}$ = 23 km/s. This provides additional evidence that the features that cause unusually blue or red color can be associated with age.

\begin{figure}
\includegraphics[width=0.5\linewidth]{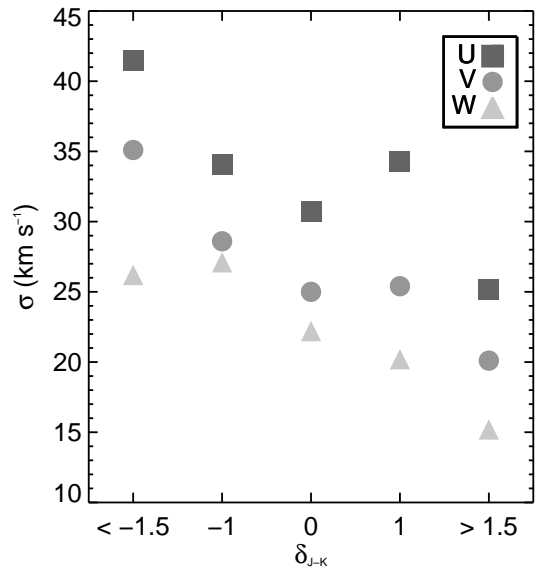} %SDSSL/kine/jkvel.pro
\caption{$UVW$ velocity dispersions for L dwarfs divided into color difference bins.} \label{fig:jksig}
\end{figure}

In order to test this, we again used a color difference ($\delta_{J-K_S}$; see section~\ref{sec:redb}) to compare the spread in color across spectral types. We separated our sample into five bins based on their color difference and then examined the kinematics in each color difference bin. There were not sufficient objects in each bin to analyze the kinematics with two components, so we found one component velocity dispersions for each bin by fitting lines to their probability plots (similar to fitting a single Gaussian function to each distribution). Figure~\ref{fig:jksig} shows the resulting $\sigma_U$,  $\sigma_V$, and $\sigma_W$ for the 5 color difference bins. It appears that the bluer L dwarfs ($\delta_{J-K_S} < -1.5$) have higher dispersions, indicating an older population.

To test whether color difference and velocity dispersion were correlated, we used Spearman's rank  test. The correlation between color difference and dispersion is significant (P $<$ 0.05) for $\sigma_V$ and $\sigma_W$, with each having a correlation coefficient of $\rho=-0.9$. The correlation between color difference and dispersion is not significant for $\sigma_U$ (P = 0.14, $\rho=-0.7$). The lack of significant correlation for $\sigma_U$ is due to the higher dispersion found in the 0.5$ > \delta_{J-K_S} > $1.5 bin, likely an effect of small number statistics. These correlations support the idea that the color difference is primarily due to an age effect. Using the relation from \citet{Weilen1977}, we derive a mean age of 4.4 Gyr for our blue outliers ($\delta_{J-K_S} < -1.5$) and 0.9 Gyr for our red outliers ($\delta_{J-K_S} > 1.5$) . We do not provide a full relation between $\delta_{J-K_S}$ and age here. Our sample is too small to derive a reliable relation, and the kinematics and colors of L dwarfs should be examined in greater detail before an explicit relation can be given. Whether the specific mechanism is the effect of clouds, gravity, metallicity, or some combination of those three properties remains to be investigated.

\section{Summary}
We identified a sample of 484 L dwarfs using a search of the SDSS spectroscopic database, 210 of which are newly discovered. Combined with previously known dwarfs, this sample has allowed us to provide updated SDSS/2MASS colors for L and T dwarfs, revealing a systematic bias in previous selection based on $J-K_S$ color criteria.

We combined photometric distance estimates, proper motions, and radial velocities to examine the three dimensional kinematics of 306 L dwarfs. The L dwarf population is best fit by models generated from two Gaussian components, suggesting that it is made up of a combination of old and young disk stars. There is a correlation between $J-K_S$ color and velocity dispersion, which confirms a suggested relationship between age and color, with younger L dwarfs having redder colors.

This correlation is especially interesting given the bias towards the selection of redder objects in previous samples of L dwarfs. The current sample of L dwarfs is likely younger than the actual population, thus explaining suggestions that the L dwarf population is younger than the local disk (as measured by the field M dwarfs). Future work is needed to ensure that the current sample of nearby dwarfs includes those with peculiar colors, and to further examine the relationship between age and color.

%%ACKNOWLEDGEMENTS
\acknowledgments
The authors would like to thank Adam Burgasser and John Bochanski for useful discussions, and the anonymous referee for helpful comments. The authors gratefully acknowledge the support of National Science Foundation (NSF) grant AST 06-07644.

This research has benefitted from the M, L, and T dwarf compendium housed at DwarfArchives.org and maintained by Chris Gelino, Davy Kirkpatrick, and Adam Burgasser. This publication makes use of data products from the Two Micron All Sky Survey, which is a joint project of the University of Massachusetts and the Infrared Processing and Analysis Center/California Institute of Technology, funded by the National Aeronautics and Space Administration and the National Science Foundation. 

Funding for the SDSS and SDSS-II has been provided by the Alfred P. Sloan Foundation, the Participating Institutions, the National Science Foundation, the U.S. Department of Energy, the National Aeronautics and Space Administration, the Japanese Monbukagakusho, the Max Planck Society, and the Higher Education Funding Council for England. The SDSS Web Site is http://www.sdss.org/.

The SDSS is managed by the Astrophysical Research Consortium for the Participating Institutions. The Participating Institutions are the American Museum of Natural History, Astrophysical Institute Potsdam, University of Basel, University of Cambridge, Case Western Reserve University, University of Chicago, Drexel University, Fermilab, the Institute for Advanced Study, the Japan Participation Group, Johns Hopkins University, the Joint Institute for Nuclear Astrophysics, the Kavli Institute for Particle Astrophysics and Cosmology, the Korean Scientist Group, the Chinese Academy of Sciences (LAMOST), Los Alamos National Laboratory, the Max-Planck-Institute for Astronomy (MPIA), the Max-Planck-Institute for Astrophysics (MPA), New Mexico State University, Ohio State University, University of Pittsburgh, University of Portsmouth, Princeton University, the United States Naval Observatory, and the University of Washington.

%%BIBLIOGRAPHY

%%FIGURES

%%TABLES

\begin{landscape}
\begin{deluxetable}{llllllllll} \tablewidth{0pt} \tabletypesize{\scriptsize}
\tablecaption{SDSS spectroscopic sample \label{tab:spec}}
\tablehead{ \colhead{SDSS Designation (SDSS J +)} & \colhead{Other Name} & \colhead{ST} & \colhead{$i$} & \colhead{$z$} & \colhead{$J$} & \colhead{$H$} & \colhead{$K_S$} & \colhead{Discovery Reference} } 
\startdata
000632.60+140606.2  &  \nodata                   &  L0  &  20.11 $\pm$ 0.03  &  18.29 $\pm$ 0.03  &  15.85 $\pm$ 0.06  &  15.30 $\pm$ 0.07  &  15.06 $\pm$ 0.15  &   1  \\
001637.61-103911.1  &  \nodata                   &  L0  &  19.73 $\pm$ 0.03  &  17.78 $\pm$ 0.02  &  15.46 $\pm$ 0.05  &  14.78 $\pm$ 0.07  &  14.53 $\pm$ 0.10  &   1  \\
001911.64+003017.6  &  \nodata                   &  L0  &  19.36 $\pm$ 0.03  &  17.57 $\pm$ 0.02  &  14.92 $\pm$ 0.04  &  14.18 $\pm$ 0.04  &  13.57 $\pm$ 0.03  &   2  \\
002209.34-011039.7  &  \nodata                   &  L0  &  19.74 $\pm$ 0.04  &  17.98 $\pm$ 0.02  &  15.82 $\pm$ 0.07  &  15.09 $\pm$ 0.08  &  14.70 $\pm$ 0.10  &   1  \\
002611.45-094341.1  &  \nodata                   &  L1  &  19.98 $\pm$ 0.03  &  18.18 $\pm$ 0.03  &  15.60 $\pm$ 0.07  &  15.00 $\pm$ 0.08  &  14.38 $\pm$ 0.07  &   3  \\
003524.43+144739.7  &  \nodata                   &  L2  &  21.12 $\pm$ 0.08  &  19.32 $\pm$ 0.08  &  \nodata  &  \nodata  &  \nodata  &   2  \\
003843.99+134339.3  &  \nodata                   &  L1  &  20.34 $\pm$ 0.05  &  18.42 $\pm$ 0.03  &  15.91 $\pm$ 0.07  &  15.20 $\pm$ 0.07  &  14.76 $\pm$ 0.10  &   2  \\
004058.93+152845.0  &  \nodata                   &  L1  &  20.62 $\pm$ 0.06  &  18.90 $\pm$ 0.04  &  16.96 $\pm$ 0.21  &  15.58 $\pm$ 0.15  &  15.60 $\pm$ 0.23  &   3  \\
\enddata
\tablerefs{(1) \citet{West2008}; (2) \citet{Hawley2002}; (3) this paper; (4) \citet{Schneider2002}; (5) \citet{Fan2000}; (6) \citet{Reid2000}; (7) \citet{Kirkpatrick2000}; (8) \citet{Zhang2009}; (9) \citet{Reid2008}; (10) \citet{Geballe2002}; (11) \citet{Cruz2003}; (12) \citet{Kirkpatrick1999}; (13) \citet{Knapp2004} (14) \citet{Martin1999}; (15) \citet{Sheppard2009}; (16) \citet{Cruz2007}; (17) \citet{Schmidt2010}; (18) \citet{Probst1984}; (19) \citet{Wilson2003}; (20) \citet{Kendall2007b}; (21) \citet{Chiu2006}}
\end{deluxetable}
\end{landscape}

\begin{deluxetable}{llllll} \tablewidth{0pt} \tabletypesize{\scriptsize}
\tablecaption{Activity \label{tab:act}}
\tablehead{ \colhead{SDSS Designation (SDSS J +)} & \colhead{Other Name} & \colhead{ST}  & \colhead{Active?} & \colhead{H$\alpha$ EW (\AA)} } 
\startdata
074227.46+412620.8 &  \nodata                   &  L0  & y & 12.93 $\pm$ 0.84 \\
074642.42+200031.8 &  2MASSI J0746425+200032    &  L1  & y & 2.44 $\pm$ 0.15 \\
081301.02+322807.9 &  \nodata                   &  L3  & n & 0.09 $\pm$ 0.00 \\
092308.70+234013.7 &  \nodata                   &  L1  & n & 3.47 $\pm$ 0.52 \\
093600.12+043147.9 &  \nodata                   &  L2  & n & -0.54 $\pm$ 1.35 \\
101707.55+130839.3 &  2MASSI J1017075+130839    &  L2  & y & 7.20 $\pm$ 0.74 \\
102204.88+020047.5 &  \nodata                   &  L0  & n & 0.44 $\pm$ 0.61 \\
102921.85+162649.8\tablenotemark{1} &  2MASSI J1029216+162652    &  L3  & y & 3.60 $\pm$ 1.05 \\
\enddata
\tablenotetext{1}{EW = 0.5\AA~\citep{Kirkpatrick2000}; EW = 1.96\AA~\citep{Reiners2008}}
\tablenotetext{2}{EW = 1.08\AA, EW = 3.19\AA~\citep{Reiners2008}}
\tablenotetext{3}{EW = 6\AA~\citep{Schmidt2007}; EW = 5.01\AA, EW = 25.65\AA~\citep{Reiners2008}}
\tablenotetext{4}{EW = 15\AA, EW $< $5\AA~(hour apart) \citep{Schmidt2007}}
\tablenotetext{5}{EW = 11\AA EW = 28\AA~(year apart)  \citep{Schmidt2007}}
\tablenotetext{6}{EW = 5\AA~\citep{Schmidt2007}; EW = 5.76\AA~\citep{Reiners2007a}}
\end{deluxetable}

\begin{deluxetable}{l | lll | lll | lll | lll | lll} \tablewidth{0pt} \tabletypesize{\tiny} %
\tablecaption{Median Colors \label{tab:col}}
\tablehead{ \colhead{}  &  
\colhead{} &  \colhead{$i-z$}  & \colhead{} & 
\colhead{} &  \colhead{$i-J$}  & \colhead{} &   
\colhead{} &  \colhead{$z-J$} &  \colhead{} &  
\colhead{} &  \colhead{$z-K_S$} & \colhead{} &  
\colhead{} &  \colhead{$J-K_S$} & \colhead{} \\
\colhead{ST} & 
\colhead{N} & \colhead{median} & \colhead{$\sigma$} &
\colhead{N} & \colhead{median} & \colhead{$\sigma$} & 
\colhead{N} & \colhead{median} & \colhead{$\sigma$} & 
\colhead{N} & \colhead{median} & \colhead{$\sigma$} &
\colhead{N} & \colhead{median} & \colhead{$\sigma$} } 
\startdata
L0 & 301 & 1.85 $\pm$ 0.01 & 0.10    & 243 & 4.23 $\pm$ 0.01 & 0.17    & 243 & 2.39 $\pm$ 0.01 & 0.14    & 238 & 3.59 $\pm$ 0.01 & 0.22    & 289 & 1.24 $\pm$ 0.01 & 0.18    \\
L1 & 100 & 1.86 $\pm$ 0.01 & 0.10    & 84  & 4.39 $\pm$ 0.02 & 0.18    & 85  & 2.51 $\pm$ 0.02 & 0.14    & 88  & 3.82 $\pm$ 0.03 & 0.26    & 135 & 1.34 $\pm$ 0.02 & 0.19    \\
L2 & 34  & 1.85 $\pm$ 0.01 & 0.08    & 29  & 4.40 $\pm$ 0.02 & 0.10    & 30  & 2.58 $\pm$ 0.03 & 0.15    & 30  & 4.04 $\pm$ 0.05 & 0.30    & 76  & 1.48 $\pm$ 0.02 & 0.20    \\
L3 & 50  & 1.88 $\pm$ 0.02 & 0.13    & 46  & 4.54 $\pm$ 0.03 & 0.23    & 46  & 2.64 $\pm$ 0.02 & 0.17    & 49  & 4.15 $\pm$ 0.05 & 0.34    & 83  & 1.60 $\pm$ 0.02 & 0.22    \\
L4 & 13  & 2.13 $\pm$ 0.05 & 0.19    & 12  & 4.97 $\pm$ 0.07 & 0.24    & 12  & 2.81 $\pm$ 0.04 & 0.14    & 14  & 4.59 $\pm$ 0.07 & 0.25    & 36  & 1.77 $\pm$ 0.05 & 0.30    \\
L5 & 10  & 2.16 $\pm$ 0.02 & 0.07    & 10  & 4.97 $\pm$ 0.04 & 0.13    & 14  & 2.82 $\pm$ 0.04 & 0.14    & 14  & 4.69 $\pm$ 0.11 & 0.43    & 42  & 1.78 $\pm$ 0.05 & 0.31    \\
L6 & 4   & 2.43 $\pm$ 0.11 & 0.23    & 3   & 5.22 $\pm$ 0.14 & 0.24    & 7   & 2.75 $\pm$ 0.04 & 0.10    & 9   & 4.25 $\pm$ 0.08 & 0.24    & 20  & 1.82 $\pm$ 0.07 & 0.29    \\
L7 & 6   & 2.59 $\pm$ 0.08 & 0.19    & 6   & 5.29 $\pm$ 0.11 & 0.26    & 9   & 2.72 $\pm$ 0.05 & 0.15    & 10  & 4.54 $\pm$ 0.08 & 0.25    & 16  & 1.77 $\pm$ 0.05 & 0.22    \\
\enddata
\end{deluxetable}

\begin{deluxetable}{l | lll | lll | lll} \tablewidth{0pt} \tabletypesize{\scriptsize}
\tablecaption{Median $J-K_S$ Colors \label{tab:JK} }
\tablehead{ \colhead{}  &  
\colhead{} &  \colhead{All Dwarfs}  & \colhead{} &  
\colhead{} &  \colhead{SDSS Spectra}  & \colhead{} &
\colhead{} &  \colhead{Dwarf Archives} & \colhead{} \\
\colhead{ST} & 
\colhead{N} & \colhead{median} & \colhead{$\sigma$} & 
\colhead{N} & \colhead{median} & \colhead{$\sigma$} & 
\colhead{N} & \colhead{median} & \colhead{$\sigma$} } 
\startdata
L0  &  289 & 1.24 $\pm$ 0.01 & 0.18 & 218 & 1.20 $\pm$ 0.01 & 0.15 & 102 & 1.33 $\pm$ 0.02 & 0.18 \\
L1  &  135 & 1.34 $\pm$ 0.02 & 0.19 & 75 & 1.29 $\pm$ 0.02 & 0.19 & 74 & 1.36 $\pm$ 0.02 & 0.19 \\
L2  &  76 & 1.48 $\pm$ 0.02 & 0.20 & 25 & 1.45 $\pm$ 0.03 & 0.15 & 59 & 1.49 $\pm$ 0.03 & 0.21 \\
L3  &  83 & 1.60 $\pm$ 0.02 & 0.22 & 40 & 1.52 $\pm$ 0.03 & 0.19 & 64 & 1.65 $\pm$ 0.03 & 0.23 \\
L4  &  36 & 1.77 $\pm$ 0.05 & 0.30 & 3 & 1.47 $\pm$ 0.10 & 0.18 & 35 & 1.77 $\pm$ 0.05 & 0.29 \\
L5  &  42 & 1.78 $\pm$ 0.05 & 0.31 & 8 & 1.82 $\pm$ 0.08 & 0.23 & 42 & 1.78 $\pm$ 0.05 & 0.31 \\
L6  &  20 & 1.82 $\pm$ 0.07 & 0.29 & 4 & 1.61 $\pm$ 0.16 & 0.32 & 19 & 1.82 $\pm$ 0.06 & 0.25 \\
L7  &  16 & 1.77 $\pm$ 0.05 & 0.22 & 3 & 1.77 $\pm$ 0.20 & 0.34 & 15 & 1.74 $\pm$ 0.06 & 0.22 \\
L8  &  15 & 1.73 $\pm$ 0.04 & 0.17 & 5 & 1.65 $\pm$ 0.09 & 0.20 & 15 & 1.73 $\pm$ 0.04 & 0.17 \\
\enddata
\end{deluxetable}

\begin{deluxetable}{llllll} \tablewidth{0pt} \tabletypesize{\scriptsize}
\tablecaption{Dwarfs with parallaxes and SDSS photometry \label{tab:plx}}
\tablehead{\colhead{SDSS coordinates} & \colhead{ST} & \colhead{M$_i$} & \colhead{$i-z$} & \colhead{$i-J$} & \colhead{plx ref}} 
\startdata
03 20 59.8 +18 54 20.9  &   M8    &   14.85  $\pm$  0.03  &  1.68  $\pm$  0.02  &  3.92  $\pm$  0.02  &  1 \\
14 35 17.2 -00 46 12.8  &  L0   &   15.32  $\pm$  1.14  &  1.78  $\pm$  0.05  &  3.87  $\pm$  0.11  &   2 \\
15 10 16.7 -02 41 07.8  &   M9    &   15.68  $\pm$  0.17  &  1.76  $\pm$  0.02  &  4.13  $\pm$  0.02  &  3 \\
22 55 29.0 -00 34 34.0  &  L0   &   15.88  $\pm$  0.35  &  1.91  $\pm$  0.03  &  4.19  $\pm$  0.06  &   2 \\
15 01 08.1 +22 50 01.7  &   M8.5    &   15.96  $\pm$  0.02  &  1.79  $\pm$  0.02  &  4.22  $\pm$  0.02  &  1 \\
14 28 43.1 +33 10 36.8  &   M9    &   15.99  $\pm$  0.03  &  1.81  $\pm$  0.02  &  4.21  $\pm$  0.02  & 4\\
14 35 35.7 -00 43 47.1  &   L3    &   16.89  $\pm$  0.78  &  1.82  $\pm$  0.07  &  4.37  $\pm$  0.13  & 2 \\ 
13 28 55.1 +21 14 46.0  &   L5    &   18.84  $\pm$  0.28  &  2.33  $\pm$  0.09  &  5.19  $\pm$  0.13  &   1 \\
17 11 45.7 +22 32 04.3  &   L6.5    &   19.75  $\pm$  0.36  &  2.33  $\pm$  0.19  &  5.06  $\pm$  0.25  & 2 \\
01 07 52.4 +00 41 56.3  &   L8    &   20.21  $\pm$  0.17  &  2.57  $\pm$  0.08  &  5.35  $\pm$  0.09  &  2 \\
08 25 19.4 +21 15 50.2  &  L7   &   20.46  $\pm$  0.06  &  2.82  $\pm$  0.06  &  5.50  $\pm$  0.06  &  1 \\
08 30 08.1 +48 28 47.3  &  L8   &   20.64  $\pm$  0.13  &  3.15  $\pm$  0.09  &  5.78  $\pm$  0.09  &  2 \\
16 32 29.2 +19 04 40.3  &   L8    &   20.77  $\pm$  0.17  &  3.11  $\pm$  0.16  &  5.81  $\pm$  0.17  &  1 \\
\enddata
\tablerefs{(1) \citet{Dahn2002}; (2) \citet{Vrba2004}; (3) \citet{Tinney1995}; (4) \citet{Monet1992}}
\end{deluxetable}

\begin{landscape}
\begin{deluxetable}{llllllllllll} \tablewidth{0pt} \tabletypesize{\tiny}
\tablecaption{Kinematics \label{tab:kine}}
\tablehead{\colhead{SDSS Designation} & \colhead{Other Name} & \colhead{Spectral} & \colhead{ST} & \colhead{$\mu_{RA}$} & \colhead{$\mu_{dec}$} & \colhead{$\mu$} & \colhead{distance} & \colhead{radial velocity} & \colhead{$U$} & \colhead{$V$} & \colhead{$W$}\\
\colhead{(SDSS J +)} & \colhead{} & \colhead{Type} & \colhead{ref} & \colhead{(mas~yr$^{-1}$)} & \colhead{(mas~yr$^{-1}$)} & \colhead{ref} & \colhead{(pc)} & \colhead{(km~s$^{-1}$)} & \colhead{(km~s$^{-1}$)} & \colhead{(km~s$^{-1}$)} & \colhead{(km~s$^{-1}$)}}
\startdata
000250.97+245413.8 &  \nodata                  &  L5.5      &   1    & 3.0 $\pm$ 23.1     & -57.8 $\pm$ 50.9   &  2         & 50.3 $\pm$ 10.4     & \nodata             & \nodata            & \nodata            & \nodata            \\
000632.60+140606.3 &  \nodata                  & L0         &   2    & 74.1 $\pm$ 44.6    & -42.2 $\pm$ 42.9   &  2         & 63.0 $\pm$ 12.0     & 26.80 $\pm$ 13.70   & -8.86 $\pm$ 13.47  & 4.58 $\pm$ 13.72   & -24.80 $\pm$ 13.51  \\
001637.62-103911.1 &  \nodata                  & L0         &   2    & -46.5 $\pm$ 63.8   & -263.1 $\pm$ 44.5  &  2         & 32.9 $\pm$ 6.3      & 25.20 $\pm$ 13.40   & 36.05 $\pm$ 10.61  & -17.24 $\pm$ 10.30  & -27.66 $\pm$ 13.10  \\
001911.64+003017.6 &  \nodata                  & L0         &   2    & -29.0 $\pm$ 18.0   & -25.0 $\pm$ 11.0   &  3         & 49.4 $\pm$ 9.4      & -4.50 $\pm$ 15.30   & 19.30 $\pm$ 4.66   & 1.86 $\pm$ 7.60    & 9.08 $\pm$ 13.49  \\
002209.33-011039.7 &  \nodata                  & L0         &   2    & \nodata            & \nodata            &  \nodata   & 66.2 $\pm$ 12.7     & 31.80 $\pm$ 17.80   & \nodata            & \nodata            & \nodata            \\
002611.45-094341.1 &  \nodata                  & L1         &   2    & \nodata            & \nodata            &  \nodata   & 61.9 $\pm$ 11.8     & -20.70 $\pm$ 3.10   & \nodata            & \nodata            & \nodata            \\
002855.49-192716.3 &  2MASS J00285545-1927165  &  L0:       &   4    & 99.5 $\pm$ 19.2    & 32.2 $\pm$ 20.9    &  2         & 23.9 $\pm$ 4.6      & \nodata             & \nodata            & \nodata            & \nodata            \\
003524.44+144739.7 &  \nodata                  & L2         &   2    & \nodata            & \nodata            &  \nodata   & 104.6 $\pm$ 20.6    & -2.40 $\pm$ 10.20   & \nodata            & \nodata            & \nodata            \\
\enddata
\tablerefs{(1) \citet{Chiu2006}; (2) this paper; (3) \citet{Jameson2008a}; (4) \citet{Reid2008}; (5) \citet{Kirkpatrick2010}; (6) H02; (7) \citet{Burgasser2006a}; (8) \citet{Kirkpatrick1999}; (9) \citet{Faherty2009}; (10) \citet{Cruz2007}; (11) \citet{Schneider2002}; (12) \citet{Cruz2009}; (13) unpublished?; (14) \citet{Kirkpatrick2000}; (15) \citet{Burgasser2003b}; (16) \citet{Knapp2004}; (17) \citet{Cruz2003}; (18) \citet{Gizis2002}; (19) \citet{Looper2007}; (20) \citet{Sheppard2009}
}
\end{deluxetable}
\end{landscape}

\begin{deluxetable}{llllllll} \tablewidth{0pt} \tabletypesize{\scriptsize}
\tablecaption{New L Dwarfs Within 30 pc \label{tab:close} }
\tablehead{\colhead{SDSS Designation} & \colhead{ST} & \colhead{$J-K_S$} & \colhead{$\delta_{J-K}$} & \colhead{distance (pc)} & \colhead{distance (pc)} & \colhead{distance (pc)} & \colhead{distance (pc)} \\
\colhead{} & \colhead{} & \colhead{} & \colhead{} & \colhead{$i-z$ vs. M$_i$}& \colhead{$i-J$ vs. M$_i$} & \colhead{ST vs. M$_J$} & \colhead{mean}} 
\startdata
SDSS J141624.09+134826.7 &  L6 & 1.03  $\pm$ 0.03 & -2.68 & 5.6 $\pm$  1.1 & 5.2 $\pm$ 0.8 & 6.7 $\pm$ 3.2 & 6.2 $\pm$ 2.0  \\
SDSS J092308.70+234013.7 &  L1 & 1.04  $\pm$ 0.03 & -1.55 & 13.7 $\pm$  2.6 & 20.7 $\pm$ 3.2 & 23.5 $\pm$ 11.0 & 21.4 $\pm$ 7.4  \\
SDSS J090023.68+253934.3 &  L7 & 1.77  $\pm$ 0.14 & 0.00 & 18.6 $\pm$  3.6 & 28.7 $\pm$ 4.8 & 25.1 $\pm$ 11.8 & 24.8 $\pm$ 8.2  \\
SDSS J135439.54+504451.0 &  L1 & 1.20  $\pm$ 0.05 & -0.75 & 22.7 $\pm$  4.3 & 28.4 $\pm$ 4.4 & 29.6 $\pm$ 13.9 & 28.0 $\pm$ 9.0  \\
SDSS J093113.23+280227.1 &  L3 & 1.25  $\pm$ 0.05 & -1.62 & 31.0 $\pm$  5.9 & 23.9 $\pm$ 3.7 & 29.0 $\pm$ 13.6 & 28.7 $\pm$ 9.1  \\
\enddata
\end{deluxetable}

\begin{deluxetable}{llll} \tablewidth{0pt} \tabletypesize{\scriptsize}
\tablecaption{Means and Dispersions of Kinematic Components \label{tab:2kin} }
\tablehead{\colhead{Velocity Vector} & \colhead{Component} & \colhead{Mean (km~s$^{-1}$)} & \colhead{$\sigma$ (km~s$^{-1}$)} } 
\startdata
$U$ & cold & 4.5 $\pm$ 0.1 & 27.2 $\pm$ 0.1 \\
$U$ & hot & $-$7.4 $\pm$ 0.4 & 72.3 $\pm$ 0.5 \\
$V$ & cold & $-$10.5 $\pm$ 0.1 & 24.5 $\pm$ 0.1\\
$V$ & hot & $-$24.4 $\pm$ 0.3 & 49.3 $\pm$ 0.4 \\
$W$ & cold & 1.2 $\pm$ 0.1 & 18.2 $\pm$ 0.1 \\
$W$ & hot & $-$7.2 $\pm$ 0.4 & 60.1 $\pm$ 0.4\\
\enddata
\end{deluxetable}

\end{document}